\documentclass[reprint,superscriptaddress,showpacs,amsmath,amssymb,aps,pra]{revtex4-1}
\usepackage{times}
\usepackage{amssymb}
\usepackage{graphicx}
\usepackage{dcolumn}
\usepackage{dsfont}
\usepackage{bm}
\usepackage{subfigure}
\usepackage[ruled]{algorithm2e}
\usepackage{hyperref}
\usepackage{appendix}

\hypersetup{colorlinks=true, citecolor=blue, urlcolor=blue, linkcolor=blue}
\begin{document}
\title{Quantum gradient descent algorithms for nonequilibrium steady states and linear algebraic systems}
\author{Jin-Min Liang}
\email{jmliang@cnu.edu.cn}
\affiliation{School of Mathematical Sciences, Capital Normal University, Beijing 100048, China}
\author{Shi-Jie Wei}
\affiliation{Beijing Academy of Quantum Information Sciences, Beijing 100193, China}
\affiliation{State Key Laboratory of Low-Dimensional Quantum Physics and Department of Physics, Tsinghua University, Beijing 100084, China}
\author{Shao-Ming Fei}
\email{feishm@cnu.edu.cn}
\affiliation{School of Mathematical Sciences, Capital Normal University, Beijing 100048, China}
\affiliation{Shenzhen Institute for Quantum Science and Engineering, Southern University of Science and Technology, Shenzhen 518055, China}
\date{Received November 12, 2021; accepted December 27, 2021; published online March 21, 2022}

\begin{abstract}
The gradient descent approach is the key ingredient in variational quantum algorithms and machine learning tasks, which is an optimization algorithm for finding a local minimum of an objective function. The quantum versions of gradient descent have been investigated and implemented in calculating molecular ground states and optimizing polynomial functions. Based on the quantum gradient descent algorithm and Choi-Jamiolkowski isomorphism, we present approaches to simulate efficiently the nonequilibrium steady states of Markovian open quantum many-body systems. Two strategies are developed to evaluate the expectation values of physical observables on the nonequilibrium steady states. Moreover, we adapt the quantum gradient descent algorithm to solve linear algebra problems including linear systems of equations and matrix-vector multiplications, by converting these algebraic problems into the simulations of closed quantum systems with well-defined Hamiltonians. Detailed examples are given to test numerically the effectiveness of the proposed algorithms for the dissipative quantum transverse Ising models and matrix-vector multiplications.

\medskip
\textbf{PACS number(s):} 02.30.Mv, 02.60.Pn, 03.67.Lx, 03.65.Yz

\medskip
\textbf{Keywords:} Quantum simulation, Quantum gradient descent algorithm, Nonequilibrium steady state, Quantum open system
\end{abstract}
\parskip=3pt

\maketitle
\section{Introduction}
Quantum computers can effectively simulate the dynamics of quantum systems \cite{Feynman1982Simulating,Bacon2010Recent,Childs2012Hamiltonian}, prepare quantum states \cite{Xin2019Preparation,Yang2019Dissipative} and solve machine learning tasks \cite{Shor1994,Grover1996,HHL2009,QML2017,Li2020Efficient,Ye2021Generic,Ran2020Tensor}, although the related algorithms generally require deep gate sequences.  In the past years, rapid advances in the construction of large-scale fault-tolerate universal quantum computers have been made based on, for instance, superconducting qubits \cite{Krantz2019A,Huang2020Superconducting,Wu2021Strong}, photos \cite{Slussarenkoa2019Photonic}, silicon quantum dots \cite{Yang2019Silicon} and ultracold trapped ions \cite{Blatt2012Quantum,Bruzewicza2019Trapped}.

Currently, we are in an era of noisy intermediate-scale quantum (NISQ) processors that may have limited resources such as a few tens to hundreds of qubits with no error correction capability, shallow circuit depth and short coherence time \cite{preskill2018quantum,larose2019variational,peruzzo2014variational,higgott2019variational,jones2019variational}. Such devices have ushered in the era of variational quantum algorithms (VQAs). VQAs aim to tackle complex problems by combining classical computers and NISQ devices \cite{benedetti2019parameterized,wang2020variational,li2021optimizing,Liu2021Hybrid,Liang2021Quantum}. Experimental evidence related to VQAs suggests that NISQ devices may improve machine learning performance for image generation \cite{Huang2021Experimental}, classification \cite{havlivcek2019supervised} and combinatorial optimization problems \cite{Harrigan2021Quantum}.

Different from VQAs, the full quantum eigensolver (FQE) finds the ground state of a hermitian Hamiltonian of a closed quantum system \cite{wei2020a,Long2006General} on a quantum computer without classical optimizers. In particular, due to the fact that the iterative optimization part utilizes the quantum gradient descent (QGD) algorithm, FQE does not require quantum-classical optimization loops like VQAs and thus can be applied totally on a quantum computer. This significant example implies that QGD as an optimization scheme is of  importance in the NISQ era \cite{Fan2021Exponential}.

Inspired by FQE for chemistry simulation, here we try to find interesting applications of QGD in open quantum systems and linear algebraic systems. In reality, the coupling to the environment is unavoidable for quantum systems and may lead to a rich variety of novel physical phenomena \cite{diehl2008quantum,verstraete2009quantum}. When the interaction with the environment is a Markovian one, the open quantum system is governed by the quantum master equation in the Lindblad form \cite{breuer2002the,rotter2009a}. As a result, the dynamics of the system give rise to incoherent characteristics such as damping and dephasing process. Such non-unitary evolution thus can not be directly implemented by normal methods designed for the isolated quantum systems \cite{rotter2009a}.

Aside from simulating open quantum systems on a quantum computer \cite{Han2021Experimental,wei2016duality}, the nonequilibrium steady state (NESS) under time-independent dissipation is also a particularly significant topic. The NESS exhibits significant properties in measurement-based quantum computation \cite{kraus2008preparation} and is topologically nontrivial \cite{diehl2011topology}. However, with the exponential growth of the Hilbert space with the number of lattice sites, exponentially many complex numbers are required to describe the full density matrix, which in practice is hard to tackle. Although some previous results based on neural networks \cite{nagy2019variational,hartmann2019variational,vicentini2019variational} and variational quantum algorithm framework \cite{yoshioka2020variational,Liu2021Variational} have been reported, the study of NESS still requires more investigation.

In this work, we provide an efficient quantum iterative algorithm for determining the NESS with the help of quantum gradient descent (QGD) algorithm. The proposed method is implemented totally on the quantum computer and thus does not require the classical-quantum optimization loop compared with previous works \cite{yoshioka2020variational,Liu2021Variational}. The quantum master equation in Lindblad form is mapped into a pure state form described by a stochastic Schr$\ddot{\textrm{o}}$dinger equation with a non-Hermitian Hamiltonian. In this case, the NESS is expressed in terms of a linear equation. We then apply the QGD algorithm  to search the ground state of the redefined hermitian Hamiltonian which involves the Liouvillian superoperator associated to the master equation. Although the output NESS is in a vector form of the density matrix, two strategies are proposed to evaluate the expectation value of physical observables for the NESS. It is worth to notice that our QGD is based on the linear combinations of unitaries \cite{Long2006General,wei2020a} instead of the quantum simulation of the gradient operator \cite{rebentrost2019quantum}. Moreover, we find broader applications of QGD in solving linear algebra problems. In particular, we solve the linear systems of equations and matrix-vector multiplications by converting these algebraic problems into hermitian Hamiltonian evolutions. Finally, we numerically test our algorithms with the dissipative quantum transverse Ising model and other toy examples.

\section{Quantum gradient descent algorithm}
As a popular optimization approach, the classical gradient descent (CGD) method plays a particularly significant role in various fields. The CGD iteratively finds a local minimum of the objective function starting from an initial guess by moving along the negative gradient of the objective function. In quantum realm, an appealing paradigm of CGD is to train a variational quantum circuit on the parameter space. These circuits run several times to optimize some objective functions which could associate to the energy of quantum states \cite{HardwareVQE2017,Parrish2019Quantum} or the loss in a quantum machine learning model \cite{Lloyd2018Quantum,Demers2018Quantum}. The first quantum version of CGD is based on the quantum simulation of the gradient operator and the quantum phase estimation \cite{rebentrost2019quantum}. Due to the substantial circuit depth, this algorithm requires more computational resources.

The second type quantum gradient descent algorithm (QGD) \cite{li2021optimizing} is based on the linear combination of unitary operators, which provides explicitly quantum circuit with amenable circuit depth and firstly demonstrates the process to optimize polynomials on a quantum simulator. Moreover, a generalized QGD gets rid of the homogeneous and even-order constraints in previous work and thus accomplishes the quantum gradient algorithm for general polynomials \cite{Gao2021Quantum}. In addition, a second-order optimization algorithm \cite{Fan2021Exponential} inspired by the classical Newton algorithm achieves a faster convergence speed compared with the above work \cite{rebentrost2019quantum,Gao2021Quantum}.

In this section, we first summarize the framework of quantum gradient descent (QGD) algorithm with the linear combination of unitary operators \cite{Gao2021Quantum,wei2020a}.

Let $f(|x\rangle):\mathbb{C}^{N}\mapsto\mathbb{R}$ be the objective function to be minimized. Given an initial quantum state $|x^{(0)}\rangle$, one iteratively updates the quantum state by using the negative gradient direction $\nabla {f(x^{(s)})}$ of the objective function. The update process is given by,
\begin{align}
|x^{(s+1)}\rangle:=|x^{(s)}\rangle-\gamma|\nabla {f(x^{(s)})}\rangle.
\end{align}
The learning rate, $\gamma>0$, determines the step length of each iteration. The gradient direction state $|\nabla {f(x^{(s)})}\rangle$ is specifically formulated by applying a gradient operator $\mathcal{G}$ (not necessary unitary) to the state $|x^{(s)}\rangle$,
\begin{align}
|x^{(s+1)}\rangle:=(I-\gamma\mathcal{G})|x^{(s)}\rangle\equiv D|x^{(s)}\rangle.
\end{align}
The non-unitary operator $D$ may be either parameter-dependent or parameter-independent in different cases. For example, in calculating the molecular ground energies and electronic structures, the non-unitary operator $D=1-2\gamma\hat{H}$ is a parameter-independent operator, where $\hat{H}$ is the Hamiltonian of the system \cite{wei2020a}.

The gradient iteration can be seen as a state evolution under the non-unitary operator $D$. This non-unitary dynamic can not be naturally simulated on quantum devices. However, it is possible to embed the non-unitary operator into a unitary operator in a larger Hilbert space. The basic idea is to denote $D$ in terms of local operators $\hat{D}[m]$. By adding the ancillary system, one performs a controlled unitary on the state $|x^{(s)}\rangle$ to prepare the state $|x^{(s+1)}\rangle$. The size of the ancillary system is determined by the number of local terms of the operator $D$.

In the following sections, we adapt QGD to prepare the nonequilibrium steady states and handle linear algebra problems, such as the linear equations with hermitian and non-Hermitian matrices and matrix-vector multiplication.

\section{Simulation of open quantum system with QGD algorithm}

\subsection{Dynamics of open quantum systems}
The dynamics of density matrix, $\rho\in\mathbb{C}^{N\times N}$, $N=2^n$, is governed by a quantum master equation in the Lindblad form \cite{lindbald1976on},
\begin{equation}\label{masterequation}
\begin{aligned}
\dot{\rho}(t)=\mathcal{L}\rho(t),
\end{aligned}
\end{equation}
where $\mathcal{L}$ is the Liouvillian superoperator such that
\begin{equation}
\begin{aligned}
\mathcal{L}\rho=-i[H,\rho]+\mathcal{D}\rho,
\end{aligned}
\end{equation}
$t$ is the time and $H$ is the Hamiltonian operator. The term $[H,\rho]$ describes the unitary part of the dynamics, while $\mathcal{D}\rho$ gives the non-unitary time evolution governed by dissipative superoperators,
\begin{equation}
\begin{aligned}
\mathcal{D}\rho=\sum_{k}\frac{\mu_k}{2}\Big[2L_k\rho L_k^{\dag}-\{L_k^{\dag}L_k,\rho\}\Big].
\end{aligned}
\end{equation}
Here each $L_k$ is the $k$th jump operator associated with a dissipative channel occurring at rate $\mu_k$. Let us consider the special but quite common case in which the Hamiltonian $H$ and the jump operator $L_k$ is usually expressed as the sum of tensor products \cite{lloyd1996universal,mahdian2020hybrid},
\begin{align}\label{localoperator}
H=\sum_{i}h_i\hat{H}[i],\,~~L_k=\sum_{j}l_j\hat{L}_{k}[j],~h_i,l_j\in\mathbb{R}.
\end{align}

Under the Choi-Jamiolkowski isomorphism \cite{havel2003robust,yoshioka2019constructing,kamakari2021digital}, $\rho=\sum_{i}p_{i}|\Psi_{i}\rangle\langle \Psi_{i}|\rightarrow|\rho\rangle=\sum_{i}p_{i}|\Psi_{i}\rangle\otimes|\Psi_{i}\rangle$, Eq. (\ref{masterequation}) is written as,
\begin{align}
|\dot{\rho}(t)\rangle=\mathcal{H}|\rho(t)\rangle
\end{align}
with the Liouville operator
\begin{align}\label{nonhermitian}
\mathcal{H}&=-i(I_{N}\otimes H^{T}-H\otimes I_{N})+\nonumber\\
&\sum_{k=1}^{n}\frac{\mu_k}{2}\Big(2L_k\otimes L_k^{*}-I_{N}\otimes L_k^{T}L_k^{*}-L_k^{\dag}L_k\otimes I_{N}\Big),
\end{align}
where $L_k^{*}$ denotes the complex conjugate of $L_k$ and $I_{N}$ is the $N\times N$ identity matrix. In this representation, the time evolution is formally solved as $|\rho(t)\rangle=e^{\mathcal{H}t}|\rho(0)\rangle$ with the initial state $|\rho(0)\rangle$.

\subsection{Searching NESS by QGD}
The NESS $\rho_{\textrm{ss}}$, $\dot{\rho}_{\textrm{ss}}=\mathcal{L}\rho_{\textrm{ss}}=0$ or $|\rho_{\textrm{ss}}\rangle=\lim_{t\rightarrow\infty}|\rho(t)\rangle$ \cite{schirmer2010stabilizing,minganti2018spectral},
corresponds to a pure state $|\rho_{\textrm{ss}}\rangle$ satisfying $\mathcal{H}|\rho_{\textrm{ss}}\rangle=0$. Since the operator (\ref{nonhermitian}) is not hermitian, we consider instead the hermitian operator $\mathcal{H}^{\dag}\mathcal{H}\succeq0$. It is straightforward to see that $|\rho_{\textrm{ss}}\rangle$ also be the ground state of $\mathcal{H}^{\dag}\mathcal{H}$. Thus, the objective function is defined as the expectation value of $\mathcal{H}^{\dag}\mathcal{H}$,
\begin{align}
f(|\rho\rangle)=\langle\rho|\mathcal{H}^{\dag}\mathcal{H}|\rho\rangle.
\end{align}

\begin{figure}[ht]
\centering
\includegraphics[width=3.3in]{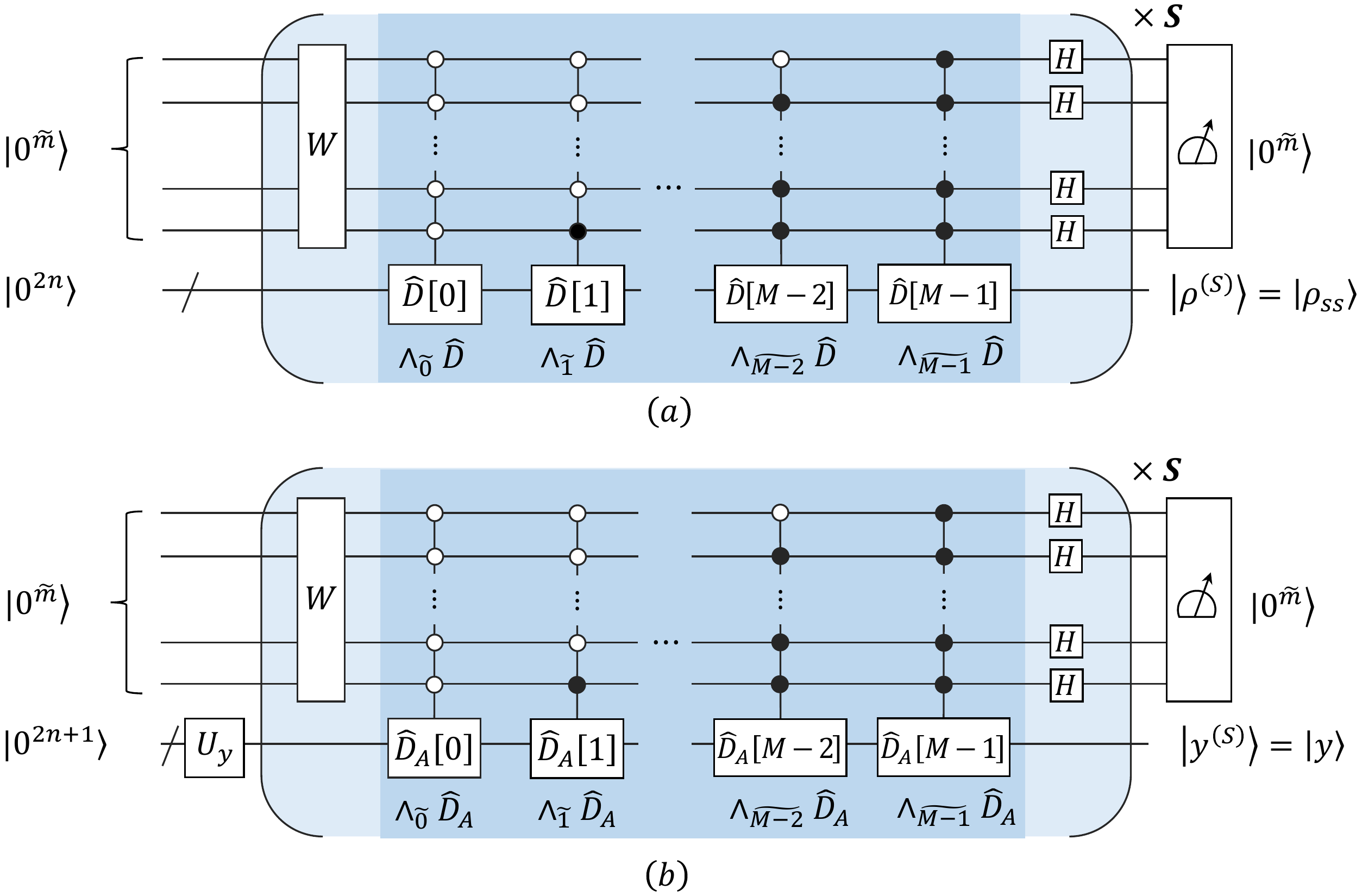}
\caption{(a) Quantum circuit for preparing the NESS. (b) Quantum circuit for solving linear equations.}
\label{circuit}
\end{figure}

We here use QGD to minimize $f(|\rho\rangle)$. The QGD iterative process can be regraded as,
\begin{align}
|\rho^{(s+1)}\rangle:&=|\rho^{(s)}\rangle-\gamma\nabla f(|\rho^{(s)}\rangle)\nonumber\\
&=|\rho^{(s)}\rangle-2\gamma\mathcal{H}^{\dag}\mathcal{H}|\rho^{(s)}\rangle=D|\rho^{(s)}\rangle.
\end{align}
Therefore, according to Eqs. (\ref{localoperator}) and (\ref{nonhermitian}), $D$ can be expressed as a sum of $M=\mathcal{O}(poly(n))$ local terms, $D=\sum_{m=0}^{M-1}d_m\hat{D}[m]\in\mathbb{C}^{N^{2}\times N^{2}}$, where the unitary $\hat{D}[m]$ consists of some local Pauli operators. After normalization, we obtain the iterative equation such that
\begin{align}
|\rho^{(s+1)}\rangle&=\frac{1}{\sqrt{C^{(s)}}}D|\rho^{(s)}\rangle\nonumber\\
&=\frac{1}{\sqrt{C^{(s)}}}\sum_{m=0}^{M-1}d_m\hat{D}[m]|\rho^{(s)}\rangle,
\end{align}
where the normalized constant,
\begin{align}
C^{(s)}&=\langle \rho^{(s)}|D^{\dag}D|\rho^{(s)}\rangle\nonumber\\
&=\sum_{m_{1},m_{2}=0}^{M-1}d_{m_{1}}d_{m_{2}}\langle \rho^{(s)}|\hat{D}^{\dag}[m_{1}]\hat{D}[m_{2}]|\rho^{(s)}\rangle,
\end{align}
can be estimated from the expectation value of $\hat{D}^{\dag}[m_{1}]\hat{D}[m_{2}]$. The Algorithm 1 below and the Fig. (\ref{circuit}.a) outline the preparation of NESS with QGD.
\begin{algorithm}[ht]
\caption{The preparation of NESS with QGD}
\LinesNumbered
\textbf{Input:} the state $|\Psi_0\rangle=|0^{\widetilde{m}}\rangle_1|\rho^{(0)}\rangle_2$ and the tolerance error $\varepsilon$.

\textbf{Output:} the state $|\rho_{\textrm{ss}}\rangle$ satisfying $|f(|\rho_{\textrm{ss}}\rangle)|\leq\varepsilon$.

\emph{Step 1.} Apply unitary $W$ on the first register and obtain the entangled state,
$$|\Psi_1\rangle=\frac{1}{\sqrt{\mathcal{N}_{D}}}\sum_{m=0}^{M-1}d_m|m\rangle_1|\rho^{(0)}\rangle_2.$$

\emph{Step 2.} Apply the controlled unitary $\mathcal{C}_{\widetilde{m}}(\hat{D})=\sum_{m=0}^{M-1}|m\rangle\langle m|\otimes\hat{D}[m]$ on both registers. We obtain
$$|\Psi_2\rangle=\frac{1}{\sqrt{\mathcal{N}_{D}}}
\sum_{m=0}^{M-1}d_m|m\rangle_1\hat{D}[m]|\rho^{(0)}\rangle_2.$$

\emph{Step 3.} Perform $\widetilde{m}$ Hadamard gates on the first register. The state of the whole system becomes
$$\begin{aligned}
|\Psi_3\rangle=\frac{1}{\sqrt{\mathcal{N}_{D}M}}|0^{\widetilde{m}}\rangle_1D|\rho^{(0)}\rangle_2.
\end{aligned}$$

\textbf{Readout.} Repeat steps 1-3 roughly $S$ times, and measure the first register. Once we obtain the state $|0^{\widetilde{m}}\rangle_1$, the output state is $|\rho_{\textrm{ss}}\rangle\approx|\rho^{(S)}\rangle$ if $|f(|\rho^{(S)}\rangle)|\leq\varepsilon$ is satisfied.
\end{algorithm}

Our QGD algorithm requires two quantum registers, the \emph{register 1} and \emph{register 2}. The input is a tensor product state $|\Psi_0\rangle=|0^{\widetilde{m}}\rangle_1|\rho^{(0)}\rangle_2$, $\widetilde{m}=\log M$. The state $|\rho^{(0)}\rangle$ is an initial guess state picked from easily prepared states such as $|\rho^{(0)}\rangle_2=|0^{2n}\rangle_2$.

In step 1, we employ the amplitude encoding method to prepare a $\widetilde{m}$-qubit input state $\sum_{m=0}^{M-1}\frac{d_m}{\sqrt{\mathcal{N}_{D}}}|m\rangle$, where $\mathcal{N}_{D}=\sum_{m=0}^{M-1}d_m^2$. The preparation takes $\mathcal{O}(poly(\widetilde{m}))$ steps if $d_m$ and $\mathcal{N}_{D}$ can be efficiently computed by classical algorithms \cite{Grover2002Creating,Soklakov2006Efficient}. The preparation process can be carried out by applying the following unitary gate on state $|0^{\widetilde{m}}\rangle$,
\begin{equation}\label{unitaryw}
\begin{aligned}
W=
\begin{bmatrix}
d_0/\sqrt{\mathcal{N}_{D}} & w_{0,1} & \cdots & w_{0,M-1}\\
d_1/\sqrt{\mathcal{N}_{D}} & w_{1,1} & \cdots & w_{1,M-1}\\
\cdots & \cdots & \cdots & \cdots\\
d_{M-1}/\sqrt{\mathcal{N}_{D}} & w_{d-1,1} & \cdots & w_{M-1,M-1}
\end{bmatrix},
\end{aligned}
\end{equation}
where the elements $\{w_{0,1},\cdots,w_{M-1,M-1}\}$ are arbitrary so long as $W$ is unitary.

In step 2, we implement the $\widetilde{m}$-qubit-controlled unitary,
\begin{align}
\mathcal{C}_{\widetilde{m}}(\hat{D})&=\prod_{m=0}^{M-1}\Bigg(|m\rangle\langle m|\otimes\hat{D}[m]+\sum_{q=0,q\neq m}^{M-1}|q\rangle\langle q|\otimes I_{N^{2}}\Bigg)\nonumber\\
&=\prod_{m=0}^{M-1}\wedge_{\widetilde{m}}\hat{D},
\end{align}
on state $|\Psi_1\rangle$ to generate the entangled state
\begin{align}
|\Psi_2\rangle=\frac{1}{\sqrt{\mathcal{N}_{D}}}
\sum_{m=0}^{M-1}d_m|m\rangle_1\hat{D}[m]|\rho^{(0)}\rangle_2.
\end{align}
The non-unitary gradient operator $D$ is realized by an extended unitary gate on a larger Hilbert space.

Without loss of generality, $\hat{D}[m]$ can be expressed as
$\hat{D}[m]=I\otimes\hat{P}_1\otimes\cdots\otimes\hat{P}_{J}\otimes I$, $\quad J<2n$,
where the Pauli operator $\hat{P}_{j}\in\{X,Y,Z\}$ acts on $j$th qubit.
The multi-qubit-controlled unitary gate $\wedge_{\widetilde{m}}\hat{D}$ can be written as
\begin{align}
\wedge_{\widetilde{m}}\hat{D}&=|m\rangle\langle m|\otimes\hat{D}[m]+\sum_{q=0,q\neq m}^{M-1}|q\rangle\langle q|\otimes I_{N^{2}}\nonumber\\
&=|m\rangle\langle m|\otimes(I\otimes\hat{P}_1\otimes\cdots\otimes\hat{P}_{J}\otimes I)\nonumber\\
&+\sum_{q=0,q\neq m}^{M-1}|q\rangle\langle q|\otimes I_{N^{2}}\nonumber\\
&=\prod_{j=1}^{J}\Bigg(|m\rangle\langle m|\otimes(I_{2^{j-1}}\otimes\hat{P}_{j}\otimes I_{2^{2n-j}})\nonumber\\
&+\sum_{q=0,q\neq m}^{M-1}|q\rangle\langle q|\otimes I_{N^{2}}\Bigg).\nonumber
\end{align}
Namely, $\wedge_{\widetilde{m}}\hat{D}$ can be decomposed into multi-qubit-controlled Pauli operators $\wedge_{\widetilde{m}}U$, $U\in\{\hat{P}_{1},\cdots,\hat{P}_{J}\}$.
Specifically, $\wedge_{\widetilde{m}}\hat{D}$ can be implemented by $\mathcal{O}(MJ)$ multiple-qubit-controlled $\textrm{SU}(2)$ unitary operators.

Last step, we apply $\widetilde{m}$ Hadamard gates on register 1 and the system state now becomes a separable one,
\begin{align}
|\Psi_3\rangle=\frac{1}{\sqrt{\mathcal{N}_{D}M}}|0^{\widetilde{m}}\rangle_1D|\rho^{(0)}\rangle_2
=\frac{1}{\sqrt{\mathcal{N}_{D}M}}|0^{\widetilde{m}}\rangle_1|\rho^{(1)}\rangle_2.\nonumber
\end{align}
Suppose that the steps 1-3 are repeated $S$ times. The final state of the system is given by
\begin{align}
|\Psi_{\textrm{final}}\rangle&=\frac{1}{(\mathcal{N}_{D}M)^{S/2}}|0^{\widetilde{m}}\rangle_1D^{S}|\rho^{(0)}\rangle_2\nonumber\\
&=\frac{1}{(\mathcal{N}_{D}M)^{S/2}}|0^{\widetilde{m}}\rangle_1|\rho^{(S)}\rangle_2.
\end{align}
When the register 1 is in state $|0^{\widetilde{m}}\rangle$, the state $|\rho^{(S)}\rangle$ could be a better approximation of the NESS $|\rho_{\textrm{ss}}\rangle$ if the $|\rho^{(S)}\rangle$ satisfies $|f(|\rho^{(S)}\rangle)|\leq\varepsilon$. Otherwise, repeat the overall procedures again until the convergence condition is satisfied. The success probability of obtaining $|0^{\widetilde{m}}\rangle$ is given by
\begin{align}
P_{\textrm{suc}}&=\langle\Psi_{\textrm{final}}|(|0^{\widetilde{m}}\rangle\langle0^{\widetilde{m}}|\otimes I_{N})|\Psi_{\textrm{final}}\rangle\nonumber\\
&=\frac{\|D^{S}|\rho^{(0)}\rangle\|^2}{(\mathcal{N}_{D}M)^{S}},
\end{align}
which decreases exponentially with respect to the number of iterations. Performing quantum amplitude amplification \cite{brassard2000quantum}, we have that the number of measurements is at most
\begin{align}
\frac{2L+1}{\max(P_{\textrm{suc}},1-P_{\textrm{suc}})}\in\Theta\Big(\frac{1}{\sqrt{P_{\textrm{suc}}}}\Big),
\end{align}
where $L=\lfloor\pi/4\theta\rfloor$ and $\sin^2\theta=P_{\textrm{suc}}$. The final success probability is at least $\max(P_{\textrm{suc}},1-P_{\textrm{suc}})$.

\subsection{Estimation of expectation values}
\begin{figure}[htbp]
\includegraphics[scale=0.6]{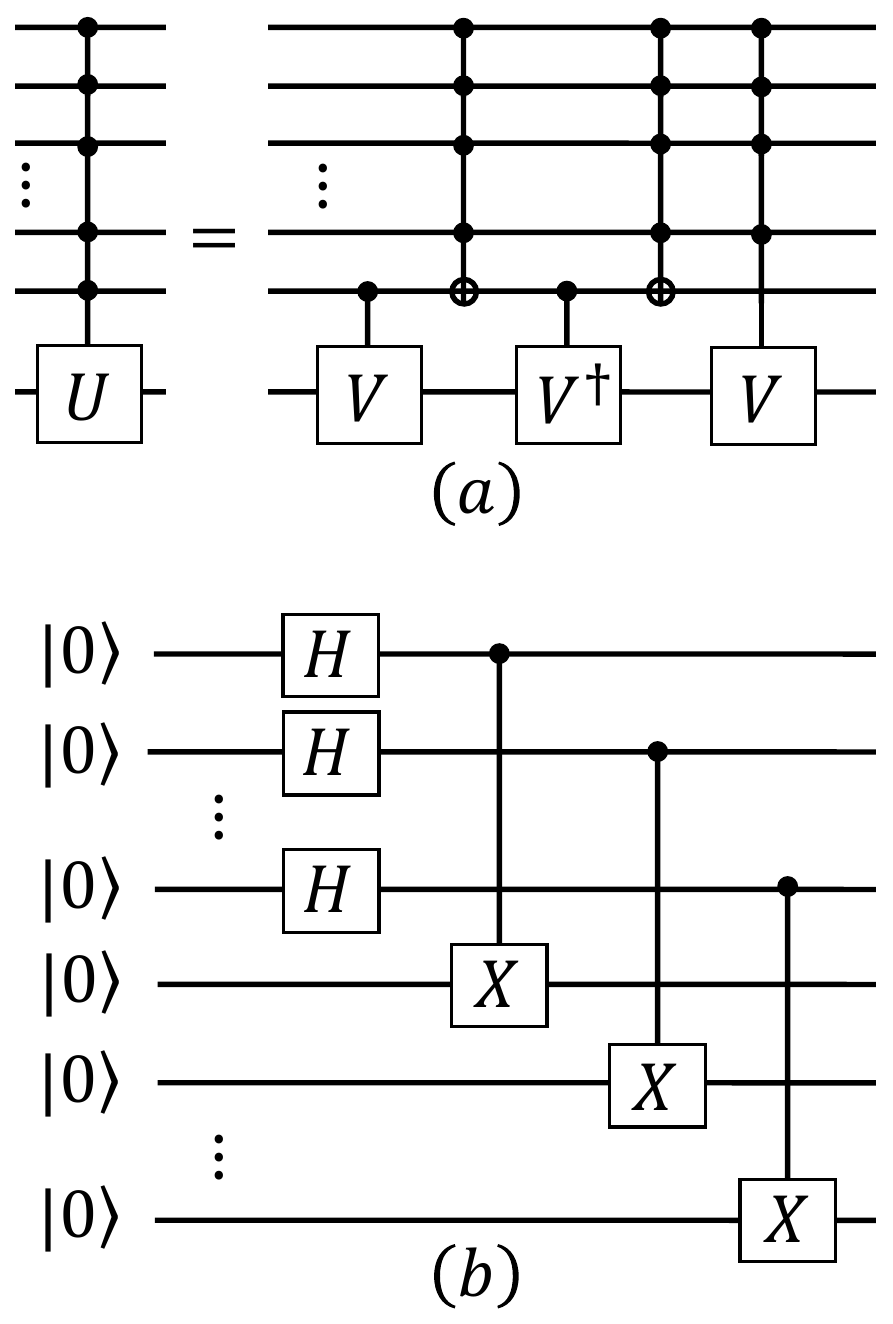}
\caption{(a) The decomposition of a controlled unitary $\wedge_{\widetilde{m}}U$ for any unitary $2\times2$ matrix $U$. $V$ is unitary and $VV=U$. (b) The unitary $U_{I_{N}}$ for preparing the state $|I_{N}\rangle$.}
\label{supplycircuit}
\end{figure}

In addition to focusing on the NESS of open quantum systems, it is also important to provide an efficient scheme to compute expectation values of physical observables. Our algorithm yields a quantum state $|\rho_{\textrm{ss}}\rangle$ related to the density matrix $\rho_{\textrm{ss}}$. Note that in our approach the state is normalized, i.e., $\langle\rho|\rho\rangle=1$. In general, the condition on the trace $\textrm{Tr}[\rho]=\langle I_{N}|\rho\rangle=1$ is not automatically fulfilled. Thus, the expectation value of an arbitrary observable $\mathcal{M}$ must be estimated as,
\begin{align}\label{expectation1}
\langle\mathcal{M}\rangle=\frac{\textrm{Tr}[\mathcal{M}\rho_{\textrm{ss}}]}{\textrm{Tr}[\rho_{\textrm{ss}}]}=
\frac{\langle I_{N}|\hat{\mathcal{M}}|\rho_{\textrm{ss}}\rangle}{\langle I_{N}|\rho_{\textrm{ss}}\rangle},
\end{align}
with respect to the density matrix $\rho_{\textrm{ss}}$ for a given observable $\mathcal{M}$. Here, the super-operator $\hat{\mathcal{M}}=\mathcal{M}\otimes I_{N}$ and the maximally entangled state
\begin{align}
|I_{N}\rangle=\sum_{i=0}^{N-1}\frac{1}{\sqrt{N}}|i\rangle|i\rangle=U_{I_{N}}|0^{2n}\rangle,
\end{align}
which is prepared by applying the unitary $U_{I_{N}}$ on an initial state $|0^{2n}\rangle$. The quantum circuit of the unitary $U_{I_{N}}$ consists of the Hadamard gate $H$ and CNOT gate, see Fig. (\ref{supplycircuit}.b.) In particular, by setting $\mathcal{M}=I_{N}$, the trace of NESS $\rho_{\textrm{ss}}$ can be calculated by $\textrm{Tr}[\rho_{\textrm{ss}}]=\langle I_{N}|\rho_{\textrm{ss}}\rangle$.

We propose two strategies to estimate the expectation value (\ref{expectation1}) from the output state of the algorithm. While the swap test calculates the square of the inner product \cite{buhrman2001quantum,fanizza2020beyond}, the magnitude and sign of the inner product are also required in our specific cases.

\emph{Strategy 1.} Our first strategy calculates the real and imaginary parts separately based on the modified swap test \cite{zhao2019quantum} or Hadamard test \cite{aharonov2008a}.

Start with an initial state $\frac{1}{\sqrt{2}}(|0\rangle+\zeta|1\rangle)$ with $\zeta=1$ or $\zeta=i$. We prepare the state $|\phi_0\rangle=\frac{1}{\sqrt{2}}(|0\rangle|I_{N}\rangle+\zeta|1\rangle|\rho_{\textrm{ss}}\rangle)$ by applying the control unitary $\wedge_{1}U$ (see Proposition 1) on the state $(1/\sqrt{2})[|0\rangle+\zeta|1\rangle]|0^{2n}\rangle$. After applying the Hadamard gate on the first qubit, we obtain the state
\begin{equation}
|\phi_1\rangle=(1/2)[|0\rangle(|I_{N}\rangle+\zeta|\rho_{\textrm{ss}}\rangle)
+|1\rangle(|I_{N}\rangle-\zeta|\rho_{\textrm{ss}}\rangle)].
\end{equation}

Now we measure the qubit system in the basis of Pauli operator $Z$. The measurement output is a random variable $\in\mathbb{Z}=\{r_{\pm}=\pm1\}$, where the output $r_{+}$ $(r_{-})$ corresponds to the measurement output state $|0\rangle$ $(|1\rangle)$. The expectation value of $\mathbb{Z}$ is given by
\begin{align}
\mathbb{E}(\mathbb{Z})&=P_{+}r_{+}+P_{-}r_{-}=2P_{+}-1\nonumber\\
&=\frac{\zeta\langle I|\rho_{\textrm{ss}}\rangle+\zeta^{*}\langle\rho_{\textrm{ss}}|I\rangle}{2}.
\end{align}
If $\zeta=1$, we have $\mathbb{E}(\mathbb{Z})=\textrm{Re}[\langle I_{N}|\rho_{\textrm{ss}}\rangle]$. If $\zeta=i$, then $\mathbb{E}(\mathbb{Z})=\textrm{Im}[\langle I|\rho_{\textrm{ss}}\rangle]$. Let $P_{+}$ be the probability of measurement outcome $r_{+}$. To estimate the probability $P_{+}$, we perform independent trials of the Bernoulli test. Then sample $R$ times and record the number $R_0$ of observed $|0\rangle$. The frequency $R_0/R$ then gives an estimator for $P_{+}$ up to a sampling error $\tilde{\varepsilon}$. From Hoeffding's inequality \cite{hoeffding1963}, we have
\begin{equation}
\begin{aligned}
P\Bigg(\Big|\frac{2R_0}{R}-1-\mathbb{E}(\mathbb{Z})\Big|\geq\widetilde{\varepsilon}\Bigg)\leq2e^{-2R\widetilde{\varepsilon}^2}=\delta.
\end{aligned}
\end{equation}
We obtain the number of measurements, $R=\mathcal{O}(\widetilde{\varepsilon}^{-2}\log \delta^{-1})$. Therefor we have the following conclusion.

\textbf{Proposition 1.} \emph{Let $\wedge_{1}U=|0\rangle\langle0|\otimes U_{I_{N}}+|1\rangle\langle1|\otimes U_{\rho_{\textrm{ss}}}$ be the controlled unitary gate with $U_{I_{N}}|0^{2n}\rangle=|I_{N}\rangle$ and $U_{\rho_{\textrm{ss}}}|0^{2n}\rangle=|\rho_{\textrm{ss}}\rangle$. There exists a quantum algorithm that calculates} $\textrm{Re}[\langle I_{N}|\rho_{\textrm{ss}}\rangle]$ \emph{and} $\textrm{Im}[\langle I_{N}|\rho_{\textrm{ss}}\rangle]$ \emph{to accuracy $\widetilde{\varepsilon}$ with success probability at least $1-\delta$ by using $\mathcal{O}(\widetilde{\varepsilon}^{-2}\log \delta^{-1})$ queries of $\wedge_{1}U$.}

Next we estimate the trace average $\textrm{Tr}[\mathcal{M}\rho]=\langle I_{N}|\hat{\mathcal{M}}|\rho\rangle$.

\textbf{Proposition 2.} \emph{Denote the hermitian operator $\mathbb{M}=X\otimes\hat{\mathcal{M}}$ and the controlled unitary gate $\wedge_{1}U=|0\rangle\langle0|\otimes U_{I_{N}}+|1\rangle\langle1|\otimes U_{\rho_{\textrm{ss}}}$ with $U_{I_{N}}|0^{2n}\rangle=|I_{N}\rangle$ and $U_{\rho_{\textrm{ss}}}|0^{2n}\rangle=|\rho_{\textrm{ss}}\rangle$. The success probability to calculate} $\langle I_{N}|\hat{\mathcal{M}}|\rho_{\textrm{ss}}\rangle$ \emph{to accuracy $\widetilde{\varepsilon}$ via measuring the expectation value of $\mathbb{M}$ on state $|\phi_0\rangle=\frac{1}{\sqrt{2}}(|0\rangle|I_{N}\rangle
+\zeta|1\rangle|\rho_{\textrm{ss}}\rangle)$ is at least $1-\delta$ by using $\mathcal{O}(\widetilde{\varepsilon}^{-2}\log \delta^{-1})$ queries of $\wedge_{1}U$.}

Proposition 2 can be proved by noting that
\begin{equation}
\textrm{Tr}[\mathbb{M}|\phi_0\rangle\langle\phi_0|]=\left\{
\begin{aligned}
&\textrm{Re}[\langle I_{N}|\hat{\mathcal{M}}|\rho_{\textrm{ss}}\rangle],~~\zeta=1, \\
&\textrm{Im}[\langle I_{N}|\hat{\mathcal{M}}|\rho_{\textrm{ss}}\rangle],~~\zeta=i
\end{aligned}
\right.
\end{equation}
due to the structure of the operator $\mathbb{M}$. Therefore, similar to the proof of Proposition 1, the number of measurements is $\mathcal{O}(\widetilde{\varepsilon}^{-2}\log \delta^{-1})$.

In the last two Propositions we have assumed that the state $|\rho_{\textrm{ss}}\rangle$ can be prepared by applying a unitary $U_{\rho_{\textrm{ss}}}$ on an initial state $|\Phi_{0}\rangle=|0^{2n}\rangle$. However, as shown in Algorithm 1, our state $|\rho_{\textrm{ss}}\rangle$ is prepared via a unitary evolution (denoted as $V$) coupling to a measurement operator $|0^{\widetilde{m}}\rangle\langle0^{\widetilde{m}}|\otimes I_{N}$. The overall process corresponds to a non-unitary process,
\begin{align}
\mathcal{E}=(|0^{\widetilde{m}}\rangle\langle0^{\widetilde{m}}|\otimes I_{N})V.
\end{align}
Thus, in order to employ the unitary $\wedge_{1}U$ introduced in the two Propositions, we need to find a unitary $U_{\rho_{\textrm{ss}}}$ to approximate this non-unitary process $\mathcal{E}$. Here, we propose a variational quantum non-unitary process approximation to find the unitary $U_{\rho_{\textrm{ss}}}$. Let $|\Phi_{0}\rangle$ be an initial system state. The evolved (normalized) state under the non-unitary operator $\mathcal{E}$ is given by $|\Phi\rangle=\frac{\mathcal{E}|\Phi_{0}\rangle}{\textrm{Tr}(\mathcal{E}|\Phi_{0}\rangle)}$. The main idea is to train a parameterized quantum circuit $\mathcal{U}(\boldsymbol\alpha)$, $\boldsymbol\alpha=(\alpha_1,\alpha_2,\cdots)$ via minimizing an objective function which quantifies the difference between the states $|\Phi\rangle$ and $\mathcal{U}(\boldsymbol\alpha)|\Phi_{0}\rangle$. $\mathcal{U}(\boldsymbol\alpha)$ consists of single-qubit rotation gates and two-qubit CNOT gates \cite{havlivcek2019supervised,commeau2020variational}.

Let us have a detailed analysis on the non-unitary approximation algorithm.
Our variational quantum non-unitary process approximation outputs a unitary $U_{\rho_{\textrm{ss}}}$ to approximately construct the non-unitary process $\mathcal{E}$ on a given initial state. A natural objective function is the square of the trace distance between $|\rho_{\textrm{ss}}\rangle$ and $|\rho(\boldsymbol\alpha)\rangle=\mathcal{U}(\boldsymbol\alpha)|\Phi_{0}\rangle$, i.e.,
$$
F(\boldsymbol\alpha)=\textrm{Tr}\Big(O\mathcal{U}^{\dag}(\boldsymbol\alpha)|
\rho_{\textrm{ss}}\rangle\langle\rho_{\textrm{ss}}|\mathcal{U}(\boldsymbol\alpha)\Big),
$$
which can be evaluated by measuring the expectation of the observable $O=I_{N^{2}}-|\Phi_{0}\rangle\langle\Phi_{0}|$ with respect to the state $\mathcal{U}(\boldsymbol\alpha)|\rho_{\textrm{ss}}\rangle$. It is straightforward to verify that the objective function can be expressed as
\begin{equation}
F(\boldsymbol\alpha)=1-|\langle\rho_{\textrm{ss}}|
\mathcal{U}(\boldsymbol\alpha)|\Phi_{0}\rangle|^2.
\end{equation}
The objective function is faithful if and only if $|\rho_{\textrm{ss}}\rangle=\mathcal{U}(\boldsymbol\alpha)|\Phi_{0}\rangle$.

The parameters $\boldsymbol\alpha$ are trained through a gradient-based optimization method. The optimal parameters can be obtained by updating $\boldsymbol\alpha=\boldsymbol\alpha-\eta\nabla F(\boldsymbol\alpha)$, where $\eta$ is the learning rate and $\nabla F(\boldsymbol\alpha)=(\partial_1F,\partial_2F,\cdots)$. The gradient information can be computed on a quantum computer via the finite difference formulae,
\begin{equation}
\partial_iF=\frac{\partial F(\boldsymbol\alpha)}{\partial \alpha_i}
=\frac{F(\boldsymbol\alpha+\Delta\alpha_i)-F(\boldsymbol\alpha-\Delta\alpha_i)}{2\Delta\alpha_i},
\end{equation}
where $\Delta\alpha_i$ is a small perturbation to $\boldsymbol\alpha$. The minimization
$\boldsymbol\alpha_{\textrm{min}}=\textrm{arg}\min_{\boldsymbol\alpha}F(\boldsymbol\alpha)$ gives rise to the optimal unitary $U_{\rho_{\textrm{ss}}}=\mathcal{U}(\boldsymbol\alpha_{\textrm{min}})$.

\emph{Strategy 2.} Our second strategy provides an directly estimation of (\ref{expectation1}) from the output $|\Psi_{\textrm{final}}\rangle$. Moreover, strategy 2 circumvents local measurement and unitary approximations, thus further reduces the computational complexity.

\textbf{Proposition 3.} \emph{Define the hermitian operator $\hat{\mathbb{M}}=X\otimes I_{M}\otimes\hat{\mathcal{M}}$ and the controlled unitary $\wedge_{1}\widetilde{U}=|0\rangle\langle0|\otimes V+|1\rangle\langle1|\otimes I_{M}\otimes U_{I_{N}}$, with} $V|0^{\widetilde{m}}\rangle|0^{2n}\rangle=|\Psi_{\textrm{final}}\rangle$ \emph{and $U_{I_{N}}|0^{2n}\rangle=|I_{N}\rangle$.}
\emph{Then we can calculate} $\langle I_{N}|\hat{\mathcal{M}}|\rho_{\textrm{ss}}\rangle$ \emph{to accuracy $\widetilde{\varepsilon}$ via measuring the expectation value of $\hat{\mathbb{M}}$ on the state}
$$|\Phi_{\textrm{final}}\rangle=\frac{1}{\sqrt{2}}
(|0\rangle|\Psi_{\textrm{final}}\rangle+\zeta|1\rangle|0^{\widetilde{m}}\rangle|I_{N}\rangle).$$
\emph{The success probability is at least $1-\delta$ by using $\mathcal{O}(\widetilde{\varepsilon}^{-2}\log \delta^{-1})$ queries of} $\wedge_{1}\widetilde{U}$.

Concerning the proof of Proposition 3 it is straightforward to check that the expectation of the observable $\hat{\mathbb{M}}$ is given by
\begin{equation}
\textrm{Tr}[\hat{\mathbb{M}}|\Phi_{\textrm{final}}\rangle\langle\Phi_{\textrm{final}}|]=\left\{
\begin{aligned}
\frac{\textrm{Re}[\langle I_{N}|\hat{\mathcal{M}}|\rho_{\textrm{ss}}\rangle]}{(\mathcal{N}_DM)^{S/2}},~~\zeta=1, \\
\frac{\textrm{Im}[\langle I_{N}|\hat{\mathcal{M}}|\rho_{\textrm{ss}}\rangle]}{(\mathcal{N}_DM)^{S/2}},~~\zeta=i.
\end{aligned}
\right.
\end{equation}
In particular, by setting $\hat{\mathcal{M}}=I_{N}$, the measurement returns to the quantity $\textrm{Tr}[\rho_{\textrm{ss}}]=\langle I_{N}|\rho_{\textrm{ss}}\rangle$.

The above strategies do not require the prior knowledge of the spectrum of  $\rho$ obtained by quantum phase estimate \cite{liang2019quantum} and quantum state tomography \cite{cramer2010efficient}. Thus our approach is experimentally friendly in near-term quantum devices.

\subsection{Computational complexity and error analysis}
In this section we present the resource complexity and error analysis. The computation complexity contains (1) the gate complexity, the number of the basic quantum gates (single and two qubit gates); (2) the qubit complexity, the number of qubits.

\textbf{Proposition 4.} \emph{Concerning the computational complexity of the preparation of $|\rho_{\textrm{ss}}\rangle$, the gate complexity of finding the NESS is $\mathcal{O}(nM\widetilde{m}^{2})$, where $M$ is the number of decomposition terms of the non-unitary gradient operator $D$ and $\widetilde{m}=\log M$. The qubit complexity is counted as $\mathcal{O}(\tilde{m}+2n)$ with $n=\log N$.}

{\sf [Proof].} The dominated source of the computational complexity is the controlled unitary $\mathcal{C}_{\widetilde{m}}(\hat{D})$ in Algorithm 1. The unitary $\mathcal{C}_{\widetilde{m}}(\hat{D})$ can be simulated by a sequence of single qubit controlled gate and Toffoli gates \cite{barenco1995elementary,Long2006Mathematical,Wen2020One,xin2017quantum,wei2018efficient}. Specifically, performing Algorithm 1 one time requires $M$ controlled unitary operator $\wedge_{\widetilde{m}}\hat{D}$. Thus the circuit depth of obtaining NESS is $\mathcal{O}(M)$. Under the Pauli decomposition of $D$, the controlled unitary operations $\wedge_{\widetilde{m}}\hat{D}$ can always be rewritten as $\mathcal{O}(n)$ multi-qubit-controlled Pauli operators $\wedge_{\widetilde{m}}U$, $U\in\{X,Y,Z\}$. A general $\wedge_{\widetilde{m}}U$ for any unitary $2\times2$ matrix $U$ can be approximately simulated by a network (Fig. \ref{supplycircuit}.a) including of unitaries $\wedge_{1}V$, $\wedge_{1}V^{\dag}$ and the Toffoli gate $T$ \cite{barenco1995elementary}. The number of basic operators scales to $\Theta(\widetilde{m}+1)$. Furthermore, the Toffoli gate $T$ on $\widetilde{m}-1$ qubits can be decomposed into $\mathcal{O}(\widetilde{m}-1)$ single-qubit and CNOT gates. The unitary operators $\wedge_{1}V$ and $\wedge_{1}V^{\dag}$ can be simulated with a cost $\mathcal{O}(1)$. Denote $\mathcal{T}_{\widetilde{m}}$ the gate complexity of decomposing $\wedge_{\widetilde{m}}U$. We have the following recurrence relation, $\mathcal{T}_{\widetilde{m}}=\mathcal{T}_{\widetilde{m}-1}+\mathcal{O}(1)+\mathcal{O}(\widetilde{m}-1)$, which implies that $\mathcal{T}_{\widetilde{m}}=\mathcal{O}(\widetilde{m}^{2})$. Hence, the total gate complexity  of simulating $\mathcal{C}_{\widetilde{m}}(\hat{D})$ is roughly
\begin{align}
\mathcal{T}_{\textrm{total}}=\mathcal{O}(nM\widetilde{m}^{2})=\mathcal{O}(\log N\cdot\log N\cdot(\log\log N)^2).\nonumber
\end{align}

As shown in Fig. (\ref{circuit}.a), the register 2 contains $2n$ qubits used to store the NESS and the state $|I_{N}\rangle$. The ancillary system requires $\widetilde{m}$ qubits determined by $M$. Thus the total qubit complexity scales to $\mathcal{O}(\widetilde{m}+2n)$. $\Box$

\textbf{Proposition 5.} \emph{Concerning the computational complexity of estimating the expectation value $\langle\mathcal{M}\rangle$, the gate and the qubit complexity are $\mathcal{O}(\textrm{poly}(n)+nM\widetilde{m}^{2})$ and $\mathcal{O}(2n)$ ($\mathcal{O}(nM\widetilde{m}^{2})$ and $\mathcal{O}(\widetilde{m}+2n)$) for strategy 1 (strategy 2), respectively.}

{\sf [Proof].}  For both strategies, it is direct to check that the qubit complexity scales to $\mathcal{O}(2n)$ and $\mathcal{O}(\widetilde{m}+2n)$. The gate complexity for strategy 1 is determined by the parameterized quantum circuit $\mathcal{U}(\boldsymbol\alpha^{*})$ which requires $\mathcal{O}(\textrm{poly}(n))$ basic quantum gates \cite{HardwareVQE2017,liang2020variational}. Therefore, by using the result of Proposition 4, the total gate complexity for strategy 1 is $\mathcal{O}(\textrm{poly}(n)+nM\widetilde{m}^{2})$. For strategy 2, we only require to perform the Algorithm 1 to prepare the state $|\Phi_{\textrm{final}}\rangle$. According to Propositions 3 and 4, we obtain the gate complexity $\mathcal{O}(nM\widetilde{m}^{2})$. $\Box$

Now we determine the upper bound of the error after $S$ time iterations. Suppose that the Algorithm 1 runs accurate enough such that
\begin{equation}\label{error}
\begin{aligned}
|f(|\widetilde{\rho}_{\textrm{ss}}\rangle)|=|\langle\widetilde{\rho}_{\textrm{ss}}|
\mathcal{H}^{\dag}\mathcal{H}|\widetilde{\rho}_{\textrm{ss}}\rangle|\leq\varepsilon.
\end{aligned}
\end{equation}
Let $|\rho_{\textrm{ss}}\rangle$ be the normalized ideal NESS and $|\widetilde{\rho}_{\textrm{ss}}\rangle=|\rho^{S}\rangle$ the actual state.

\textbf{Proposition 6.} \emph{Denote $\tau_1^{2}=|\langle\phi_1|\rho^{(0)}\rangle|^2$ the overlap between an initial state $|\rho^{(0)}\rangle$ and the state $|\phi_1\rangle$ is the largest eigenstate of operator $D$. The approximation error $\varepsilon$ has an upper bound $$\frac{\kappa(1-\tau_1^2)}{\tau_1^2}\Big(\frac{\delta_2}{\delta_1}\Big)^{2S}$$ scaling with $\tau_1^2$, iterations $S$, the gap between largest and lowest eigenvalues of $\mathcal{H}^{\dag}\mathcal{H}$, $\kappa=\lambda_{\textrm{max}}-\lambda_1$. $\delta_1$ and $\delta_2$ are the two largest eigenstates of the operator $D$.}

{\sf [Proof].}  Consider the eigenvalue decomposition of the hermitian matrix $D=\sum_{r=1}^{N^2}\delta_r|\phi_r\rangle\langle\phi_r|$ with the eigenvalues arranged in order, $\delta_1\geq\delta_2\geq\cdots\geq\delta_{N^2}=\delta_{\textrm{min}}\geq0$. The eigenvalues of $\mathcal{H}^{\dag}\mathcal{H}$ are then $\lambda_r=(1-\delta_r)/2\gamma$, $\lambda_1\leq\lambda_2\leq\cdots\leq\lambda_{N^2}=\lambda_{\textrm{max}}$. We express the initial state $|\rho^{(0)}\rangle$ in the basis of the eigenvectors $\{|\phi_r\rangle\}$,
\begin{equation}
|\rho^{(0)}\rangle=\sum_{r=1}^{N^2}\tau_r|\phi_r\rangle,~~~\sum_{r}|\tau_r|^2=1,
\end{equation}
with $\tau_r$ some coefficients. After applying the non-unitary operator $D$ to $|\rho^{(0)}\rangle$ $S$ times, we have
\begin{align}
D^{S}|\rho^{(0)}\rangle&=\sum_{r=1}^{N^2}\tau_r\delta_r^{S}|\phi_r\rangle\nonumber\\
&=\delta_1^{S}\sum_{r=1}^{N^2}\tau_r\Big(\frac{\delta_{r}}{\delta_{1}}\Big)^{S}|\phi_r\rangle.
\end{align}
As $\lim_{S\rightarrow\infty}(\delta_{r}/\delta_{1})^{S}=0$ for all $r$, for larger $S$ we can approximate the ground state of $\mathcal{H}^{\dag}\mathcal{H}$ as
\begin{equation}
|\rho_{ss}\rangle\approx|\widetilde{\rho}_{ss}\rangle=|\rho^{S}\rangle=\frac{1}{\sqrt{C^{(S)}}}D^{S}|\rho^{(0)}\rangle,
\end{equation}
where the normalization constant
\begin{align}
C^{(S)}=\langle\rho^{(0)}|D^{S\dag}D^{S}|\rho^{(0)}\rangle
=\delta_1^{2S}\sum_{r=1}^{N^2}\tau_r^2\Big(\frac{\delta_{r}}{\delta_{1}}\Big)^{2S}|\phi_r\rangle.
\end{align}

The error satisfies
\begin{align}
\varepsilon&=|f(|\widetilde{\rho}_{ss}\rangle)-f(|\rho_{ss}\rangle)|\nonumber\\
&=\Bigg|\frac{\sum_{r=2}^{N^2}\tau_{r}^2(\lambda_{r}-\lambda_{1})(\delta_r/\delta_1)^{2S}}{\sum_{r=1}^{N^2}\tau_{r}^2(\delta_r/\delta_1)^{2S}}\Bigg|\nonumber\\
&\leq\kappa\Bigg|\frac{\sum_{r=2}^{N^2}\tau_{r}^2(\delta_r/\delta_1)^{2S}}{\sum_{r=1}^{N^2}\tau_{r}^2(\delta_r/\delta_1)^{2S}}\Bigg|\nonumber\\
&\leq\frac{\kappa}{\tau_1^2}\sum_{r=2}^{N^2}\tau_{r}^2(\delta_r/\delta_1)^{2S}\nonumber\\
&\leq\frac{\kappa}{\tau_1^2}(\delta_2/\delta_1)^{2S}\sum_{r=2}^{N^2}\tau_{r}^2\nonumber\\
&=\frac{\kappa(1-\tau_1^2)}{\tau_1^2}\Bigg(\frac{\delta_2}{\delta_1}\Bigg)^{2S}.
\end{align}
Therefore, the approximation error $\varepsilon$ has an upper bound $\mathcal{O}(\frac{\kappa(1-\tau_1^2)}{\tau_1^2}(\delta_2/\delta_1)^{2S})$ scaling with $S$, $\tau_1^2$ and $\kappa=\lambda_{\textrm{max}}-\lambda_1$. $\Box$

The above error bound shows that the convergence is geometric with ratio $\delta_2/\delta_1$. This implies that the Algorithm 1 would converge slowly if an eigenvalue is close to the dominant eigenvalue of the operator $D$. Hence, one should choose an the initial state $|\rho^{(0)}\rangle$ with a larger overlap $|\langle\phi_1|\rho^{(0)}\rangle|^2$.

The estimation error of the expectation value $\langle\mathcal{M}\rangle$ after $S$ time iterations is bounded by a function $g(\varepsilon)$,
\begin{align}
E&=|\widetilde{\langle\mathcal{M}\rangle}-\langle\mathcal{M}\rangle|\nonumber\\
&=\Bigg|\frac{\langle I|\hat{\mathcal{M}}|\widetilde{\rho}_{\textrm{ss}}\rangle}{\langle I|\widetilde{\rho}_{\textrm{ss}}\rangle}-
\frac{\langle I|\hat{\mathcal{M}}|\rho_{\textrm{ss}}\rangle}{\langle I|\rho_{\textrm{ss}}\rangle}\Bigg|\leq g(\varepsilon),
\end{align}
where $\widetilde{\langle\mathcal{M}\rangle}$ denotes the approximation result of $\langle\mathcal{M}\rangle$. Clearly $g(\varepsilon)$ goes to zero if $\varepsilon=0$.

\section{QGD for linear algebra problems}
In this section, we focus on implementing the QGD algorithm to tackle the linear algebra problems such as solving linear equations and preparing the matrix-vector multiplication states. The core of our strategy is to transform these problems to the simulation of a well-defined Hamiltonian system.

\subsection{QGD for linear equations}
Given a sparse matrix $A\in\mathbb{R}^{N\times N}$ and a state vector $|b\rangle$. The problem is to solve the linear equation $A|x\rangle=|b\rangle$. Here, if $A$ is not hermitian, one can always transform the problem into an equivalent one with hermitian matrix $\tilde{A}=|0\rangle\langle1|\otimes A+|1\rangle\langle0|\otimes A^\dagger$. The solution can be expressed as $|x\rangle=\frac{A^{-1}|b\rangle}{\|A^{-1}|b\rangle\|}$. We define the Hamiltonian \cite{Subasi2019Quantum},
\begin{align}
H_{A}=(X\otimes A)(I_{2N}-|+,b\rangle\langle+,b|)(X\otimes A),
\end{align}
where the state $|+,b\rangle=\frac{1}{\sqrt{2}}(|0\rangle+|1\rangle)\otimes|b\rangle$. It is easy to check that $|+,x\rangle$ is the ground state of the Hamiltonian $H_{A}$ with ground energy $0$, $H_{A}(|+\rangle\otimes A^{-1}|b\rangle)=H_{A}|+,x\rangle=0$. It is worthwhile to notice that the Hamiltonian $H_A$ has another different form, $H_A=A^{\dag}(I-|b\rangle\langle b|)A$, as suggested in \cite{Xu2020Variational,LiuWu2021Variational}. It can be verified that the solution $|x\rangle$ is the unique eigenstate associated with the minimum eigenvalue zero of $H_A$.

Concerning the quantum setting in solving the linear equations, let $A=\sum_{i=0}^{K_A-1}a_iA[i]$ be an efficient Pauli decomposition of the matrix $A$, where $a_i>0$, $K_A=\mathcal{O}(poly(n))$, $n=\log N$ and $A[i]$ are Pauli strings.
For classically access to the right-hand side vector we also assume that $|b\rangle\langle b|=\sum_{i=0}^{K_b-1}b_ib[i]$ is an efficient Pauli decomposition
of the hermitian matrix $|b\rangle\langle b|$, with $b_i>0$, $K_b=\mathcal{O}(poly(n))$ and $b[i]$ the Pauli strings.
The Hamiltonian $H_A$ can then be efficiently encoded,
\begin{align}
H_{A}=(X\otimes A)(I_{2N}-\frac{X+I_2}{2}\otimes|b\rangle\langle b|)(X\otimes A).
\end{align}

In order to search the ground state of Hamiltonian $H_A$, we define the objective function as
\begin{equation}
f(y)=\langle y|H_{A}|y\rangle.
\end{equation}
It is clear that the minimum of $f(y)$ is the desired result with respect to $|y\rangle=|+\rangle\otimes|x\rangle$. The gradient of the objective function is given by
\begin{equation}
\nabla f(y)=2H_{A}|y\rangle=\mathcal{G}_{A}|y\rangle.
\end{equation}
Thus the gradient descent iteration can be regraded as the evolution of $|y^{(t)}\rangle$ under the operator $H_{A}$,
\begin{align}
|y^{(s+1)}\rangle&:=|y^{(s)}\rangle-\gamma\nabla {f(|y^{(s)}\rangle)}=D_{A}|y^{(s)}\rangle,
\end{align}
where $\gamma$ is the learning rate, $D_{A}=I_{2N}-\gamma\mathcal{G}_{A}$. Hence, the gradient descent iteration can be also viewed as an evolution of $|y^{(s)}\rangle$ under the non-unitary operator $D_{A}$. $D_{A}$ consists of a polynomial number (with respect to the number of qubits) of tensor products of local Pauli strings, $D_{A}=\sum_{m=0}^{M-1}d_m\hat{D}_{A}[m]$. $D_{A}$ cannot be implemented directly in quantum circuits as it is not unitary. Such non-unitary evolution can be realized in unitary quantum circuits by adding additional qubits \cite{Long2006General,Cao2010Restricted}. We use the following four steps to achieve the iterative process, see Fig. (\ref{circuit}.b).

\emph{Step 1. Prepare the initial state}. The register contains a work system and an ancillary system. At first, one picks an initial vector $y^{(0)}=(y_0^{(0)},y_1^{(0)},\cdots,y_{2N-1}^{(0)})^T$. The $(n+1)$-qubit quantum input state $|y^{(0)}\rangle=\sum_{i=0}^{2N-1}\frac{y_{i}^{(0)}}{\|y^{(0)}\|}|i\rangle$ can be prepared by, for instance, employing the amplitude encoding. This encoding uses $(n+1)$-qubit to represent vector $y^{(0)}$ and requires resources that are polynomial in the number of qubits, $\mathcal{O}(poly(n+1))$ if the norm of $\|y^{(0)}\|$ can be efficiently computed \cite{Grover2002Creating,Soklakov2006Efficient}. Note that the exact decomposition of universal gate preparing a general state $|y^{(0)}\rangle$ requires the CNOT gate complexity $\mathcal{O}(N)$ \cite{Bergholm2005Quantum,Plesch2011Quantum}. Another state preparation approach is the quantum random access memory which consumes $\mathcal{O}(poly(n))$ queries to the oracle that access the element of $y^{(0)}$ \cite{Giovannetti2008Architectures}.

In our iteration algorithm, we only need an easily prepared quantum state, i.e., tensor product state $|y^{(0)}\rangle=|0\cdots0\rangle$.
Then we add $\widetilde{m}=\log M$ qubits and prepare the state $\sum_{m=0}^{M-1}\frac{d_m}{\sqrt{\mathcal{N}_{D_{A}}}}|m\rangle$ by applying the unitary $W$ given in (\ref{unitaryw}), where $\mathcal{N}_{D_{A}}=\sum_{m=0}^{M-1}d_m^2$ is the normalization constant. The state of the whole system now is
\begin{equation}
\begin{aligned}
|\Psi_0\rangle=\sum_{m=0}^{M-1}\frac{d_m}{\sqrt{\mathcal{N}_{D_{A}}}}|m\rangle|y^{(0)}\rangle.
\end{aligned}
\end{equation}

\emph{Step 2. Implement the non-unitary operator}. We apply a series of ancillary controlled operators,
\begin{align}
\mathcal{C}_{\widetilde{m}}(\hat{D}_A)&=\sum_{m=0}^{M-1}|m\rangle\langle m|\otimes\hat{D}_{A}[m]=\prod_{m=0}^{M-1}\wedge_{\widetilde{m}}\hat{D}_A,
\end{align}
on the whole state space, where $\wedge_{\widetilde{m}}\hat{D}_A=|m\rangle\langle m|\otimes\hat{D}_{A}[m]+\sum_{q=0,q\neq m}^{M-1}|q\rangle\langle q|\otimes I_{N^{2}}$. The whole system state is transformed into
\begin{align}
|\Psi_1\rangle&=\mathcal{C}_{\widetilde{m}}(\hat{D}_A)|\Psi_0\rangle\nonumber\\
&=\sum_{m=0}^{M-1}\frac{d_m}{\sqrt{\mathcal{N}_{D_{A}}}}|m\rangle\otimes\hat{D}_{A}[m]|y^{(0)}\rangle.
\end{align}
The work qubits and the ancilla qubits are now entangled.

\emph{Step 3. Combine the states from different subspaces}. We perform $\widetilde{m}$ Hadamard gates on the ancillary register to combine the states $\hat{D}_{A}[m]|y^{(0)}\rangle$. The system state becomes
\begin{align}
|\Psi_2\rangle&=\frac{1}{\sqrt{\mathcal{N}_{D_{A}}M}}|0^{\widetilde{m}}\rangle\sum_{i=0}^{M-1}d_m\hat{D}_{A}[m]|y^{(0)}\rangle\nonumber\\
&=\frac{1}{\sqrt{\mathcal{N}_{D_{A}}M}}|0^{\widetilde{m}}\rangle D_{A}|y^{(0)}\rangle\nonumber\\
&=\frac{1}{\sqrt{\mathcal{N}_{D_{A}}M}}|0^{\widetilde{m}}\rangle|y^{(1)}\rangle.
\end{align}

\emph{Step 4. Measurement}. After repeating the above three steps $S$ times, the final state of the system is of the form,
\begin{align}
|\Psi_{\textrm{final}}\rangle&=\frac{1}{(\mathcal{N}_{D_{A}}M)^{S/2}}|0^{\widetilde{m}}\rangle D_{A}^{S}|y^{(0)}\rangle\nonumber\\
&=\frac{1}{(\mathcal{N}_{D_{A}}M)^{S/2}}|0^{\widetilde{m}}\rangle|y^{(S)}\rangle.
\end{align}
Then we measure the former $\widetilde{m}$ qubits and obtain the classical bits string $\{0\cdots0\}$. The system state collapses to $|y^{(S)}\rangle$ which is an approximate solution of $|y\rangle$ if $|f(y^{(S)})|\leq\varepsilon$. Otherwise, we repeat the overall procedures again until the convergence condition is satisfied. The success probability of obtaining the state $|0^{\widetilde{m}}\rangle$ is given by
\begin{align}
P_{\textrm{suc}}&=\langle\Psi_{\textrm{final}}|(|0^{\widetilde{m}}\rangle\langle0^{\widetilde{m}}|\otimes I_{N})|\Psi_{\textrm{final}}\rangle\nonumber\\
&=\frac{\|D_{A}^{S}|y^{(0)}\rangle\|^2}{(\mathcal{N}_{D_{A}}M)^{S}},
\end{align}
which decreases exponentially with respect to the number of iteration steps. Using amplitude amplification \cite{brassard2000quantum}, we have that $\mathcal{O}(P_{\textrm{suc}}^{-1/2})$ measurements are sufficient. Clearly, the qubit complexity is $\mathcal{O}(n+1+\widetilde{m})$.

\subsection{QGD for matrix-vector multiplication}
Given an hermitian matrix $A\in\mathbb{R}^{N\times N}$ and an state $|b\rangle$, our goal is to prepare the normalized state,
\begin{equation}
|y\rangle=\frac{A|b\rangle}{\|A|b\rangle\|}.
\end{equation}
Defining the Hamiltonian $\widetilde{H}_{A}=I_{N}-\frac{A|b\rangle\langle b|A^{\dag}}{\|A|b\rangle\|^2}$ \cite{Xu2020Variational}, one can easily check that the ground state of $\widetilde{H}_{A}$ with ground energy $0$ is the state $|y\rangle$. The expectation value $\langle\Phi|\widetilde{H}_{A}|\Phi\rangle$ of $\widetilde{H}_{A}$ with respect to an arbitrary state $|\Phi\rangle$ satisfies
\begin{align}
\langle\Phi|\widetilde{H}_{A}|\Phi\rangle&=\langle\Phi|\Big(I_{N}-\frac{A|b\rangle\langle b|A^{\dag}}{\|A|b\rangle\|^2}\Big)|\Phi\rangle\\
&=1-\frac{|\langle\Phi|A|b\rangle|^2}{\|A|b\rangle\|^2}\geq0,
\end{align}
where the equality holds if and only if state $|\Phi\rangle$ is the ground eigenstate $|y\rangle$ of $\widetilde{H}_{A}$. Hence, we define the objection function,
\begin{equation}
f(y)=\langle y|\widetilde{H}_{A}|y\rangle.
\end{equation}
The gradient descent iteration can be regraded as an evolution of $|y^{(s)}\rangle$ under the operator $\tilde{H}_{A}$,
\begin{align}
|y^{(s+1)}\rangle&:=|y^{(s)}\rangle-\gamma\nabla {f(|y^{(s)}\rangle)}=\widetilde{D}_A|y^{(s)}\rangle,
\end{align}
where the non-unitary operator $\widetilde{D}_A=I_{N}-2\gamma\widetilde{H}_{A}$. We only need to substitute the Hamiltonian $H_{A}$ with $\widetilde{H}_{A}$ and apply the algorithm discussed in last subsection to find the matrix-vector multiplication state $|y\rangle$. But, here, we require $n=\log N$ qubits to store and prepare $|y\rangle$ rather than $n+1$ qubits.

In order to estimate the normalization constant
\begin{align}\label{constant}
\|A|b\rangle\|^2=\langle b|A^{\dag}A|b\rangle,
\end{align}
we assume availability of an efficient $n$-qubit quantum circuit $U_{b}$ such that $U_{b}|0^{n}\rangle=|b\rangle$ \cite{Huang2019near}.
The constant (\ref{constant}) is written as
\begin{align}
\|A|b\rangle\|^2&=\sum_{m}|a_m|^2\langle b|A^{\dag}[m]A[m]|b\rangle\nonumber\\
&=\sum_{m}|a_m|^2\langle0^{n}|U_{b}^{\dag}A^{\dag}[m]A[m]U_{b}|0^{n}\rangle,
\end{align}
where each term can be evaluated via the Hadamard test \cite{aharonov2008a} or the measurement on the Pauli basis.

\section{Numerical experiments}
In this section we present some examples to illustrate our approaches.

\subsection{The dissipative quantum transverse Ising model}
The Hamiltonian of the dissipative quantum transverse Ising model is given by
\begin{equation}\label{hamiltonian}
\begin{aligned}
H=\frac{J}{4}\sum_{\langle k_1,k_2\rangle=1}^nZ^{(k_1)}Z^{(k_2)}+\frac{h}{2}\sum_{k=1}^{n}X^{(k)},
\end{aligned}
\end{equation}
where $X^{(k)}$ $(Z^{(k)})$ are the Pauli operator $X$ $(Z)$ acting on the $k$-th qubit. The first term denotes the nearest-neighbor spin-spin interaction depending only on the $z$ components with coupling strength $J$. The second term accounts for a local and uniform magnetic field along the transverse direction $x$. The site-dependent jump operator (the Lindbald operator) is $L_k=\sigma_{+}^{(k)}$, $L_k^{T}=L_k^{\dag}=\sigma_{-}^{(k)}$ and $L_k^{*}=L_k$, where $\sigma_{\pm}^{(k)}=(X^{(k)}\pm iY^{(k)})/2$.
Substitute Eq. (\ref{hamiltonian}) into Eq. (\ref{nonhermitian}), we have for $n=2$,
$$
\begin{aligned}
\mathcal{H}=&-i\Big[\frac{J}{4}\Big(Z^{(3)}Z^{(4)}-Z^{(1)}Z^{(2)}\Big)+\frac{h}{2}X^{(3412)}\Big]\\
&+\frac{\mu_1}{2}\Big(2\sigma_{+}^{(1)}\sigma_{+}^{(3)}-\sigma_{-}^{(3)}\sigma_{+}^{(3)}-\sigma_{-}^{(1)}\sigma_{+}^{(1)}\Big)\\
&+\frac{\mu_2}{2}\Big(2\sigma_{+}^{(2)}\sigma_{+}^{(4)}-\sigma_{-}^{(4)}\sigma_{+}^{(4)}-\sigma_{-}^{(2)}\sigma_{+}^{(2)}\Big),
\end{aligned}
$$
where $X^{(3412)}=X^{(3)}+X^{(4)}-X^{(1)}-X^{(2)}$.

\begin{figure}[ht]
\centering
\subfigure[]{\includegraphics[width=3in]{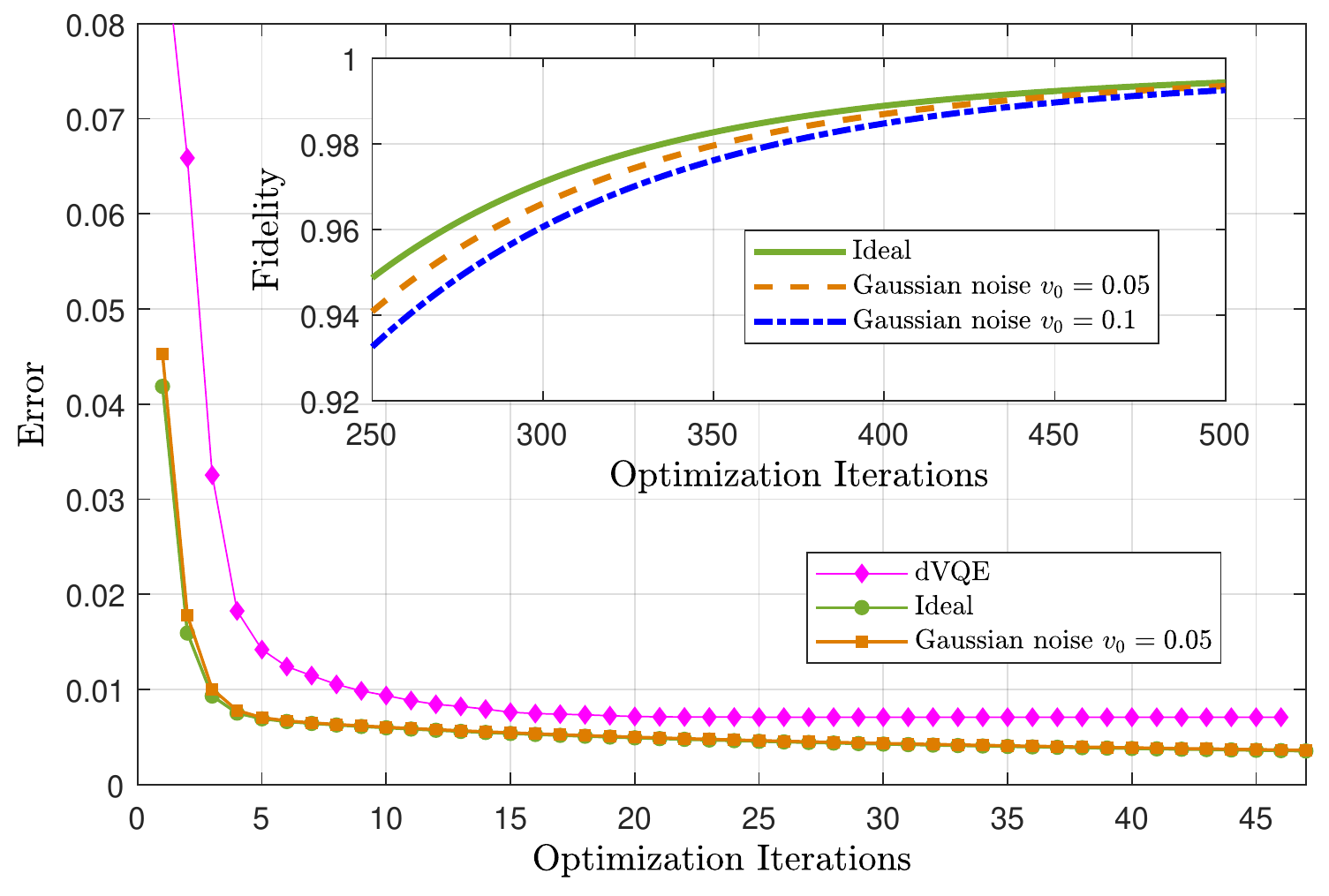}}
\subfigure[]{\includegraphics[width=3in]{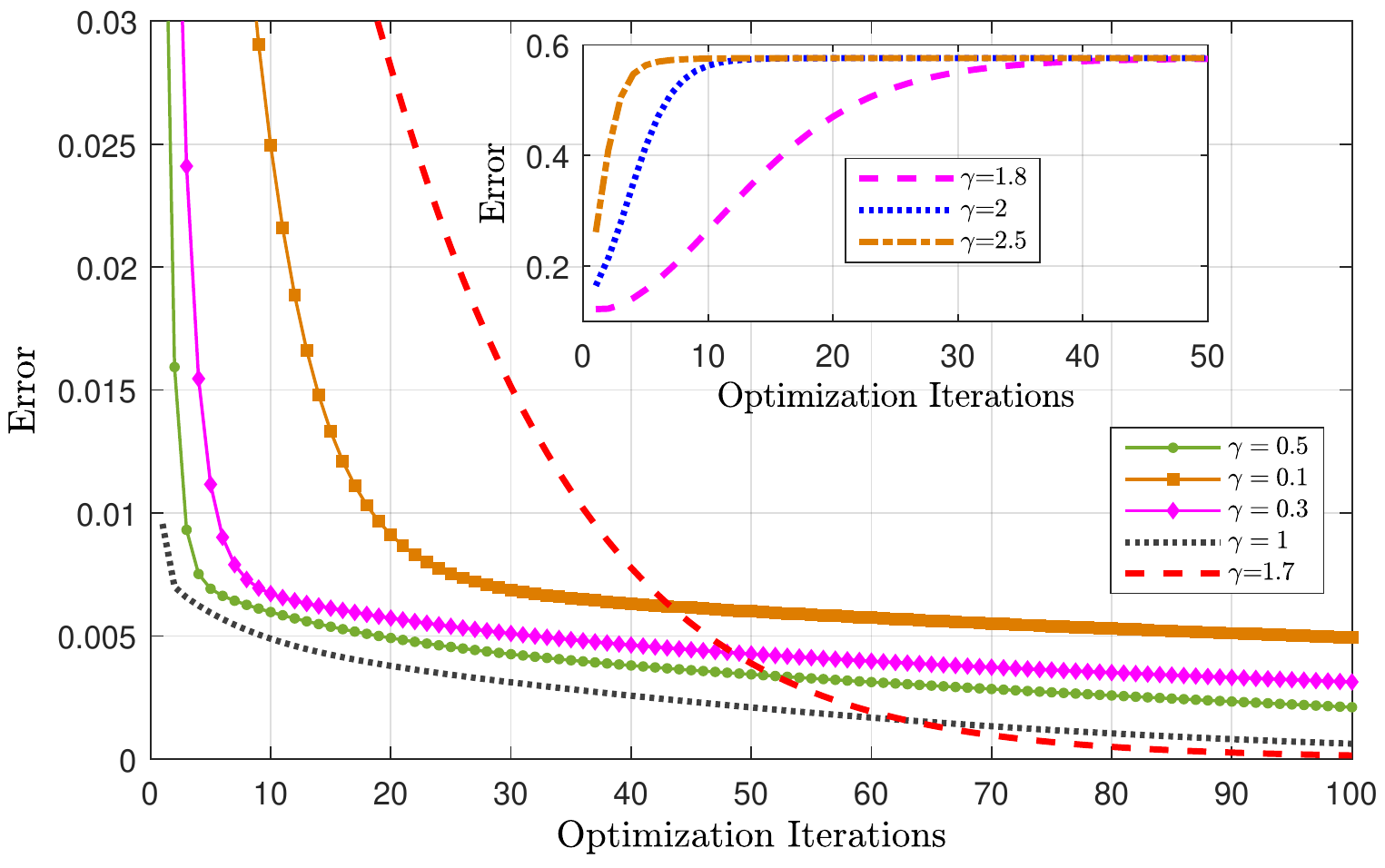}}
\caption{Numerical results of dissipative quantum transverse Ising model. (a) Error and fidelity as functions of the optimization iteration with and without noise channel. The learning rate $\gamma=0.5$. (b) Different learning rates $\gamma=0.1,0.3,0.5,1,1.7$.}
\label{result1}
\end{figure}

We set $J=h=1$, $\mu_1=\mu_2=0.1$, and the initial state $|\rho^{(0)}\rangle=|+\rangle^{\otimes4}$ with the learning rate $\gamma=0.5$, where $|+\rangle=\frac{1}{\sqrt{2}}(|0\rangle+|1\rangle)$. Because of the unavoidable errors in the operations, the non-unitary operator $D$ should be substituted by the noise dynamical map $\mathcal{E}(D,v_0)=D+v_0I_{N^2}$. Consider the noise model modeled by a Gaussian noise with variance $\sigma^{2}=1$ and mean $\omega=0$, $v_0\rightarrow\mathcal{N}(\omega,\sigma^{2})$. The result error is given by $\varepsilon=f(|\rho^{(s)}\rangle)$.
In Fig. 3 we compare the numerical results of that with and without Gaussian noise. When the optimization iterations increase, the error $\varepsilon$ converges to the minimal value zero in both situations. It is clear that the noise error converges to zero slower than the ideal error. In particular, $\varepsilon=3.57\times10^{-3}$ and $\varepsilon=3.65\times10^{-3}$ for ideal and Gaussian noise ($v_0=0.05$), respectively, with iteration $S=500$. Another quantity to evaluate our algorithm is the fidelity $F=|\langle\rho^{(S)}|M|\rho_{\textrm{ss}}\rangle|^2$. The inset of in Fig. (3.a) illustrates that the ideal fidelity is $0.99434$ which is higher than the Gaussian noise fidelities $0.99349$ ($v_0=0.05$) and $0.99253$ ($v_0=0.1$), at iteration $S=500$. Thus, the numerically behavior ensures that our algorithms are robust to Gaussian noise.

Furthermore, we compare our results with the dissipative-system variational quantum eigensolver (dVQE) \cite{yoshioka2020variational}. In dVQE, the optimization of the loss function $G(\boldsymbol\alpha)=\langle\rho^{(0)}|U^{\dag}(\boldsymbol\alpha)\mathcal{H}^{\dag}\mathcal{H}U(\boldsymbol\alpha)|\rho^{(0)}\rangle$ starts with a same state $|\rho^{(0)}\rangle$. The parameterized quantum circuit
\begin{align}
U(\boldsymbol\alpha,\boldsymbol\beta)=\Big[\mathcal{V}(\boldsymbol\beta)\otimes\mathcal{V}^{*}(\boldsymbol\beta)\Big]
\prod_{n=1}^{2}\textrm{CNOT}_{n,n+2}\Big[\mathcal{U}(\boldsymbol\alpha)\otimes I_{4}\Big],\nonumber
\end{align}
where
\begin{align}
\mathcal{U}(\boldsymbol\alpha)=&\Big[R_y(\alpha_1)\otimes R_y(\alpha_2)\Big]\mathcal{C}_{1}R_y(\alpha_3)\times\nonumber\\
&\Big[R_y(\alpha_4)\otimes R_y(\alpha_5)\Big]\mathcal{C}_{1}R_y(\alpha_6)
\end{align}
and
\begin{align}
\mathcal{V}(\boldsymbol\beta)=&\Big[R_y(\beta_1)R_x(\beta_2)\otimes R_y(\beta_3)R_x(\beta_4)\Big]\mathcal{C}_{1}Z\times\nonumber\\
&\Big[R_y(\beta_5)R_x(\beta_6)\otimes R_y(\beta_7)R_x(\beta_8)\Big]\mathcal{C}_{1}Z\times\nonumber\\
&\Big[R_y(\beta_9)R_x(\beta_{10})\otimes R_y(\beta_{11})R_x(\beta_{12})\Big].
\end{align}
The rotation operator $R_y(\alpha_i)=e^{-i\alpha_i/2Y}$, $\mathcal{C}_1Z=|0\rangle\langle0|\otimes I_2+|0\rangle\langle0|\otimes Z$, and $\mathcal{V}^{*}(\boldsymbol\beta)$ denotes the complex conjugate of $\mathcal{V}(\boldsymbol\beta)$. We apply the classical gradient descent algorithm to optimize the $18$ variational parameters. As shown in Fig. (\ref{result1}.a), our results converge faster than dVQE. The minimal error of dVQE is $\varepsilon=7.082\times10^{-3}$ compared with our results $3.57\times10^{-3}$ (ideal result) or $3.65\times10^{-3}$ (Gaussian noise, $v_{0}=0.05$) at step $S=46$.

Finally, we perform numerical simulation under different learning rates. As shown in Fig. (\ref{result1}.b), the result error does not converge to the minimal value when $\gamma>1.7$. So we should choose the proper learning rate in practical situation.

\subsection{Example for matrix-vector multiplication}
We now consider the linear equation $A|x\rangle=|b\rangle$ with
\begin{align}
A=0.9Z^{(1)}Z^{(2)}+0.3692X^{(2)}+0.1112X^{(1)},
\end{align}
and $|b\rangle=|0^{n}\rangle$. Let us consider the case of $n=3$ and the initial state $|x^{(0)}\rangle=|+^{n}\rangle$ with learning rate $\gamma=0.3$, where $|+\rangle=\frac{1}{\sqrt{2}}(|0\rangle+|1\rangle)$. The defined Hamiltonian $H_A$ is then encoded in a $4$-qubit system. As $|b\rangle\langle b|=\frac{1}{8}(X+Z)^{3}$, $H_A$ can be expressed as $H_{A}=(X\otimes A)(I_{2N}-\frac{1}{16}(X+I)\otimes(X+Z)^{3})(X\otimes A)$. As a result, the non-unitary operator $D_A$ also has similarly an Pauli decomposition.

\begin{figure}[ht]
\centering
\includegraphics[width=3in]{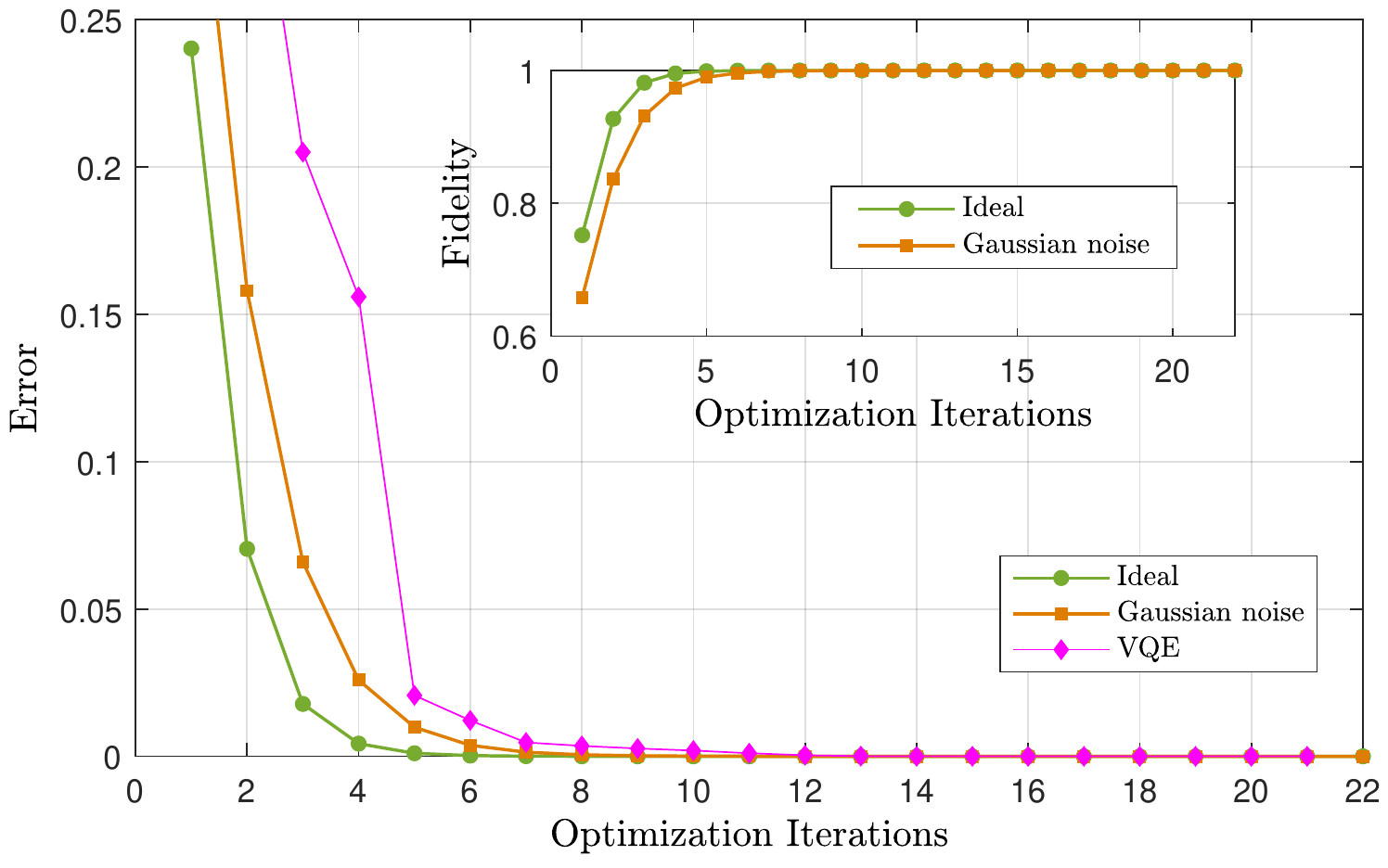}
\caption{Numerical results of matrix-vector multiplication. The relations among the error $\varepsilon$, the fidelity $F$ and the number of iterations.}
\label{result2}
\end{figure}

With the growth of the number of iterations, Fig. (\ref{result2}) shows that the error $\varepsilon$ of our approach can converges to the local minimum of the objective function. In particular, the fidelity $F=0.999$ and the error $\varepsilon=1.4475\times10^{-15}$ when $S=20$. Even at the presence of Gaussian noise ($v_{0}=0.5$), our result still approaches the ideal value, $\varepsilon=4.585\times10^{-9}$ and $F=0.9999$ when $S=20$. For a comparison we calculate the ground state of $H_{A}$ by using VQE \cite{peruzzo2014variational}. As shown in Fig. (\ref{result2}), our result converges faster than that obtained from the VQE.

\section{Conclusion and Discussion}
In conclusion, based on the quantum gradient descent algorithm we have presented a quantum approach for the (nondegenerate) nonequilibrium steady state problem of open quantum systems. Our method is scalable since the gate and the qubit complexity is polynomial functions of the size of open quantum systems. Capturing the characteristic property of NESS, we introduced two strategies to evaluate the expectation value of the interested physical observable associated with the output state given by vector forms of the density matrix $\rho_{\textrm{ss}}$. Furthermore, we have adapted the quantum gradient descent algorithm to solve linear algebra problems including linear equation and matrix-vector multiplications. The basic idea is to convert the linear algebra problems into the simulation of a closed quantum system with a well-defined Hamiltonian. Our quantum linear solver may also serve as a subroutine in solving the Poisson equations with Dirichlet boundary conditions \cite{LiuWu2021Variational}. On the one hand, the whole optimization procedures of our methods are implemented on a quantum computer. On the other hand, our approach does not need to perform measurements for the expectation values of the Hamiltonian, neither the classical optimization feedbacks, compared with the recent works \cite{yoshioka2020variational,Liu2021Variational}. Thus the computational complexity is substantially reduced. Furthermore, different from that the VQAs optimize the cost function in the parameter space, our algorithms optimize the cost function on the original state space. With the increase of the iteration, our algorithm can always obtain a better approximation of the NESS or ground state. Thus another advantage of our strategies is that the barren plateaus do not appear in the quantum gradient descent procedure.

This work focused on the first-order optimization method, but the second-order optimization method (the Newton method) \cite{Li2021Quantum} may also be used in the preparation of NESS. Aside from the preparation of NESS, it is possible to simulate the dynamics of quantum open systems on a double-dimension Hilbert space. Our approaches highlight the fact that a non-unitary operator composing of local unitary terms may be implemented on NISQ devices. Motivated by the idea of quantum imaginary time evolution \cite{Mcardle2019Variational,Motta2020Determining}, approximate simulation of non-unitary evolutions may be possible with a parameterized quantum circuit. Once the optimal parameters are learned, the non-unitary evolution can be carried out by a low-depth quantum circuit which is amenable for NISQ devices. One of the advantages of this strategy is that no any other ancillary systems are required compared with the existing schemes.

\bigskip

Acknowledgements: We would like to thank Dong Liu and Qiao-Qiao Lv for their helpful suggestions and discussions. This work is supported by NSF of China (Grant No. 12075159, 12171044), Beijing Natural Science Foundation (Z190005), Academy for Multidisciplinary Studies, Capital Normal University, the Academician Innovation Platform of Hainan Province, and Shenzhen Institute for Quantum Science and Engineering, Southern University of Science and Technology (No. SIQSE202001). The National Natural Science Foundation of China under Grants No. 12005015.

\end{document}